\documentclass[iop,revtex4]{emulateapj}
\usepackage{epstopdf}
\usepackage{graphics,graphicx}
\usepackage{helvet}
\usepackage[utf8]{inputenc}
\usepackage[T1]{fontenc}
\usepackage{soul}
\usepackage{hyperref}
\usepackage{amsmath, amssymb, gensymb}
\usepackage{ulem}
\usepackage{natbib}
\usepackage{microtype}
\usepackage{multirow}
\usepackage{morefloats}
\usepackage{rotating}
\usepackage{xspace}
\usepackage{color}
\usepackage{xspace}
\newcommand{\sod} 		{SO$_2$}
\newcommand{\Tsod}	{$^{34}$SO$_2$}

\newcommand{\Tso}		{$^{34}$SO}
\newcommand{\htcop}	{H$^{13}$CO$^+$}
\newcommand{\meta}	{CH$_3$OH}

\newcommand{\hh}		{HH~80-81}
\newcommand{\stI}		{{\it I}}
\newcommand{\stQ}		{{\it Q}}
\newcommand{\stU}		{{\it U}}
\newcommand{\kms}		{~km~s$^{-1}$}
\newcommand{\mJy}		{~mJy~beam$^{-1}$}

\newcommand{\msun}	{~M$_{\sun}$\xspace}
\newcommand{\lsun}		{~L$_{\sun}$\xspace}

\newcommand{\cmt}		{~cm$^{-3}$\xspace}
\newcommand{\vlsr}		{~$v_{\rm LSR}$\xspace}

\newcommand{\eg}    	{e.\,g.,}
\newcommand{\Mm}		{$\pm$}
\newcommand{\Area}	{$\int I_\nu^\mathrm{MM1}\,dv$}

\begin{document}

\title{The circumestellar disk of the B0 protostar powering the \hh\ radio jet}

\author{J.M. Girart\altaffilmark{1}}
\author{R. Estalella\altaffilmark{2}}
\author{M. Fern\'andez-L\'opez\altaffilmark{3}}
\author{S. Curiel\altaffilmark{4}}
\author{P Frau\altaffilmark{1}}
\author{R. Galvan-Madrid\altaffilmark{5}}
\author{R. Rao\altaffilmark{6}}
\author{G. Busquet\altaffilmark{1}}
\author{C. Ju\'arez\altaffilmark{1}}

\altaffiltext{1}{Institut de Ci\`encies de l'Espai (IEEC-CSIC), Can Magrans, S/N, 08193 Cerdanyola del Vall\`es, Catalonia, Spain}
\altaffiltext{2}{Departament de F{\'\i}sica Qu\`antica i Astrof{\'\i}sica, Institut de Ci\`encies del Cosmos (ICC), Universitat de Barcelona (IEEC-UB), Mart{\'\i} Franqu\`ees 1, 08028 Barcelona, Catalonia, Spain}
\altaffiltext{3}{Instituto Argentino de Radioastronom{\'\i}a, CCT-La Plata (CONICET), C.C.5, 1894, Villa Elisa, Argentina}
\altaffiltext{4}{Instituto de Astronom{\'\i}a, Universidad Nacional Aut\'onoma de M\'exico (UNAM), Apartado Postal 70-264, 04510 M\'exico, DF, M\'exico}
\altaffiltext{5}{Instituto de Radioastronom\'{\i}a y Astrof\'{\i}sica (UNAM), 58089 Morelia, M\'exico}
\altaffiltext{6}{Institute of Astronomy and Astrophysics, Academia Sinica, 645 N. Aohoku Place, Hilo, HI 96720, USA}

\begin{abstract}
We present  subarcsecond angular resolution observations carried out with the Submillimeter Array (SMA) at 880~$\mu$m centered at the B0-type protostar GGD27~MM1, the driving  source of the parsec scale \hh\ jet. We constrain its polarized continuum emission to be $\lesssim0.8\%$ at this wavelength. Its submm spectrum is dominated by sulfur-bearing species tracing a rotating disk--like structure (SO and SO$_2$ isotopologues mainly), but also shows HCN-bearing and \meta\ lines, which trace the disk and the outflow cavity walls excavated by the \hh\ jet. The presence of many sulfurated lines could indicate the presence of shocked gas at the disk's centrifugal barrier or that MM1 is a hot core at an evolved stage.  The resolved SO$_2$ emission traces very well the disk kinematics and we fit the SMA observations using a thin-disk Keplerian model, which gives the inclination (47$\degr$), the inner ($\lesssim170$~AU) and outer ($\sim950-1300$~AU) radii and the disk's rotation velocity (3.4\kms\ at a putative radius of 1700~AU). We roughly estimate a protostellar dynamical mass of 4-18\msun.  MM2 and WMC cores show, comparatively, an almost empty spectra suggesting that they are associated with extended emission detected in previous low-angular resolution observations, and therefore indicating youth (MM2) or the presence of a less massive object (WMC).
\end{abstract}


\section{Introduction}\label{intro}

It is well stablished that most low-mass stars develop accreting disks in their earliest stages previous to the pre-main sequence phase (\citealt{1987Adams, 1987Shu, 1991Butner}; for the first HST images of disks see \citealt{1993Odell, 1994Odell, 1996Burrows}).  These disks may be the seed where planets are formed (see recent reviews by \citealt{2014Helled} and \citealt{2014Baruteau}). The properties of the accretion disks are better studied in the T Tauri stage since the surrounding material has already infallen into the disk--protostar system or has been dispersed by the strong outflow activity \citep[e.g.,][]{1998Hogerheijde, 2004Andre}. Still recent works have found disks even in the earliest embedded stages of Class 0 protostars \citep[e.g.,][]{2012Tobin, 2013Murillo, 2014Rao}.  Class 0 disks are, in general, also smaller in size ($\la$50~AU; e.g., \citealt{2013Tobin, Segura-Cox16})  than T Tauri disks (200-300~AU; e.g., \citealt{2012Takakuwa, 2014Harsono, 2014Pietu}).  Disks are usually detected using their (sub)mm continuum emission and the emission of major CO isotopes. From the spectral line observations, the kinematical structure of disks around T Tauri protostars is found consistent with Keplerian motions \citep{1998Guilloteau, 1999Guilloteau, 2007Brinch, 2008Lommen, 2009Schaefer, 2009Jorgensen, 2012Takakuwa, 2014Pietu}, while it is more difficult to disentangle the kinematic signature of rotation and infalling motions in disks/envelopes of Class 0 protostars \citep{2007Takakuwa,2009Brinch,2012Tobin}.
  
Accretion disks around massive stars are a much rarer phenomenon. This can be explained due to an observational bias:  massive young stars dynamically evolve much faster and are located at farther distances. However, it is still a matter of debate whether the most massive stars are ever associated with accretion disks. Most of the reported accretion disks around massive stars are mostly associated with early B-type protostars \citep{2005Cesaroni,2005Patel,2009FrancoHernandez,FL11a,2012Wang,2013SanchezMonge,Beltran16}, although there are some evidence of disks around O-type stars \citep{Johnston15, Ilee16}.

The massive star-forming region GGD27 (IRAS 18162$-$2048) is located at a distance of 1.7~kpc \citep{Gyulbudaghian78}.  It harbors a cluster of IR, mm and cm sources tracing young stellar objects in different evolutionary stages \citep{Aspin91, Aspin92, Gomez95, Stecklum97, Gomez03, Qiu08, Qiu09}. Some of these sources are associated with molecular outflows and jets. In particular, \citet{FL13} find three molecular outflows powered by protostars embedded in MM2; two of them appear to be monopolar. In addition, the main mm source, MM1, launches the spectacular (14~pc long) and highly collimated radio jet  known as HH~80-81-80N,  which is powered by a massive early B-type protostar \citep{Marti93, Marti95, Marti98, Masque12, Masque15}.  The HH~80-81-80N radio jet is the first one where polarized emission due to relativistic electrons has been detected indicating the presence of a magnetic field  aligned with the jet  \citep{Carrasco10}. This is highly indicative that the jet is launched from an accretion disk \citep{2014Frank}.  Indeed, recent (sub)millimeter high-angular resolution interferometric observations show evidence of the presence of an accretion disk surrounding the massive protostar \citep[][hereafter FL11a, FL11b and CG12]{FL11a, FL11b, CG12}:  a hot ($T\simeq150$~K) and very dense ($n(\mathrm{H}_2)\simeq 10^9$~\cmt) molecular rotating disk-like structure with a radius of $\sim 0\farcs5$ (850 AU), with its rotation axes parallel to the radio jet, and surrounding a dusty disk with a radius of $\sim 0\farcs15$ (200~AU; FL11a, CG12).  The estimated centrifugal radius, $\simeq 650$~AU (CG12), suggests that within the uncertainties the observed molecular and dusty rotating structure is an accretion disk. From the available observations, a dynamical mass (disk plus protostar) of 11--15~M$_\odot$ and an accretion rate $\sim 10^{-4}$~M$_\odot$~yr$^{-1}$  (FL11b, CG12) have been inferred.

In this paper we present SMA observations carried out at 345~GHz at an angular resolution of $0\farcs4$.  The observations are briefly described in \S~\ref{obs} and the molecular line and dust continuum maps are presented in \S~\ref{res}. An analysis of the kinematic behavior of the rotating disk-like structure from a selected sample of molecular line channel maps is shown in \S~\ref{ana}. In \S~\ref{dis} we discuss the chemical and physical properties of the molecular gas detected with the SMA.  Finally, in \S~\ref{con} we draw our main conclusions.

\section{Observations}\label{obs}

The SMA observations were taken on 2011 July 18 and October 3 in the extended and very extended configurations, respectively.  The observations were done using the 345~GHz receivers and the quarter wave plates (QWP).  During the observations the QWP are rotated in order to measure the four circular polarization cross correlations for each baseline,  LL, RR, RL and RL (R and L stands for right and left circular polarization).  \citet{Marrone06} and \citet{Marrone08} describe in detail the SMA polarimeter as well as the polarization calibration strategy.    The receiver was tuned to cover the 332.1-336.0 and 344.0-347.9~GHz frequency ranges in the lower (LSB) and upper  (USB) sidebands, respectively. The phase center of the telescope was pointing at $\alpha$(J2000.0)$=18^{\rm h}19^{\rm m}12\fs10$ and  $\delta$(J2000.0)$=$-$20\degr47\arcmin 30\farcs00$.  The correlator provided a spectral resolution of about 0.8~MHz (i.e., 0.7~km~s$^{-1}$ at 345~GHz).  The gain calibrator was the QSO J1733$-$130. The bandpass and polarization calibrator was 3c454.3, which was observed in a parallactic angle range of $\sim 120\arcdeg$. The absolute flux scale was determined from observations of Callisto. The flux uncertainty was estimated to be $\sim20$\%.  The instrumental polarization was  corrected at an accuracy level of $\simeq$ 0.1\%. The data were reduced using the MIRIAD software package \citep{Wright93}. The continuum and line molecular emission were separated in the visibility space using the MIRIAD task {\it uvlin}.  An iterative process of phase-only self-calibration was performed using the continuum Stokes $I$ data. We started with a time interval of 20 minutes for the gain solutions. Then we decreased the interval in the subsequent steps until we reached an interval of 5 minutes.  At the end of the self-calibration the $rms$ noise decreased by about 30\%. The derived gain solutions were applied to the molecular line data. 

The continuum emission maps were obtained using the whole available bandwidth by setting a robust weighting of 0.5 and selecting  the multi-frequency synthesis option in the MIRIAD task {\it invert}.  The resulting synthesized beam full width at half maximum (FWHM) was $0\farcs48\times0\farcs35$ with a position angle of $29\degr$. The achieved $rms$ noise  for the Stokes \stI\  continuum map was 3.4~\mJy, and for the Stokes \stQ\ and \stU\ continuum maps was 1.6~\mJy.   The larger $rms$ noise of the Stokes \stI\ is due to the limited dynamic range of the array. For the molecular line emission, channel maps were done using a robust weighting of 2.0 (equivalent to Natural weighting). This yielded synthesized beams sizes (FWHM, PA) between $0\farcs45\times0\farcs34$, $31\arcdeg$ and $0\farcs47\times0\farcs35$, $31\arcdeg$, which are the values for the lines with the highest and lowest frequencies, respectively.  Figures were created using the GREG package (from the GILDAS\footnote{GILDAS data reduction package is available at http://www.iram.fr/IRAMFR/GILDAS}  software). The CO 3-2 maps from the observations presented here were already presented in \citet{FL13}.

%
%

\begin{figure}[h]
\includegraphics[width=\columnwidth]{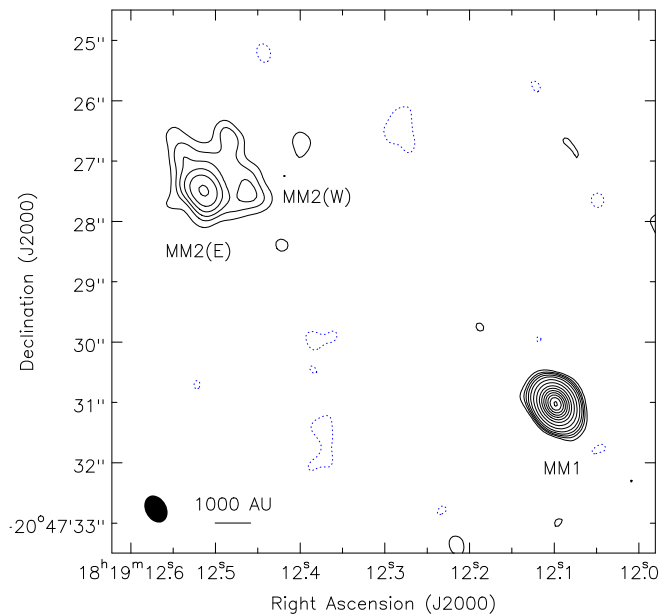}
\caption{SMA 880~$\mu$m continuum contour map of the GGD27 region. Contours are
$-3$, 3, 5, 7, 10, 15, 25, 40, 60, 80, 100, 120, 140, 160 and 180 times the $rms$ noise, 3.4\mJy. The synthesized beam, $0\farcs48\times0\farcs35$ with a position angle of $29\degr$, 
is shown in the bottom--left corner of the image. The labels indicate the source names. The cross marks the position of the Warm Molecular Core \citep{Qiu09}.
} 
\label{Fcont}
\end{figure}

\begin{table}[htp]
\begin{center}
\caption{880$\mu$m  Continuum emission \label{T1}}
\begin{tabular}{lcccc}
\hline
		& R.A. & DEC. & $I_\nu^\mathrm{peak}$ & $S_\nu^\mathrm{int}$ \\
Source 	& 18$^{\rm h}$19$^{\rm h}$ & $-20^\circ 47'$ & (mJy~beam$^{-1}$) & (mJy) \\
\hline
MM1 	& $12\fs098$ & $31\farcs02$ & 624\Mm4 & 732\Mm9 \\
MM2(E) 	& $12\fs515$ & $27\farcs50$ & 102\Mm4 & 350\Mm23$^a$  \\
MM2(W) 	& $12\fs462$ & $27\farcs50$ &   33\Mm4 & --  \\
WMC$^b$	& $12\fs450$ & $24\farcs70$& $\la 16$ \\
\hline
\end{tabular}
\\
$^a$ Total flux of MM2(E) and MM2(W) \\
$^b$ Position from \citep{Qiu09}
\end{center}
\end{table}

\begin{figure*}[h]
\includegraphics[width=18.0cm]{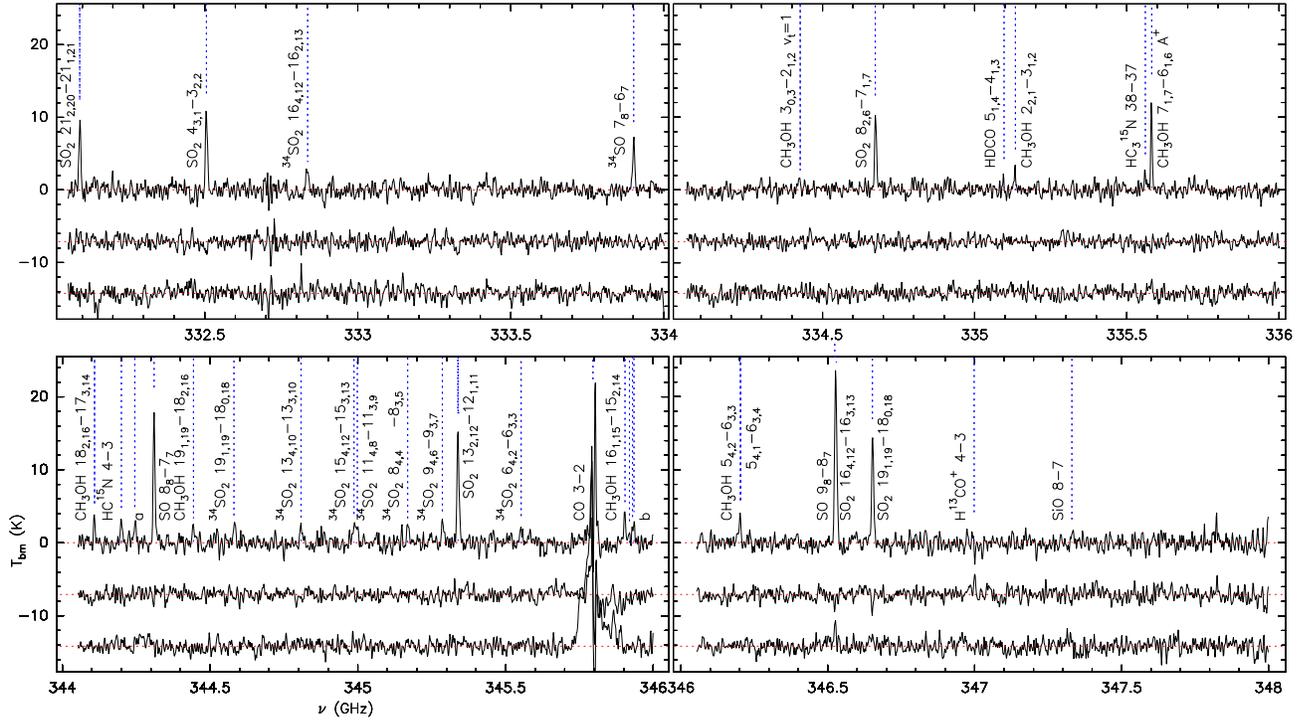}
\caption{Spectra toward GGD 27 over the observed bandwidth. The four panels 
show the different frequency ranges covered by the LSB and USB. For each panel 
the top, middle and bottom spectra are taken at the MM 1, MM 2 and WMC positions. 
All these spectra were averaged over a region of about 
2~arcsec$^2$. The conversion factor from Jy to K is $\sim 65$. 
The line with an "a" label  is the  $^{34}$SO$_2$ 10$_{4,6}$-10$_{3,7}$.  The lines with a "b" label are \meta\ 18$_{-3,16}$-17$_{-4,14}$,  $^{34}$SO$_2$ 17$_{4,14}$-17$_{3,15}$ and 
 OS$^{17}$O 15$_{4,11}$-15$_{3,12}$. 
}
\label{Fspec}
\end{figure*}

\begin{table}[htp]
\begin{center}
\caption{Molecular line parameters\label{T2}}
\begin{tabular}{ccrrr}
\hline
Molecular 			& $\nu$		& E$_l$	& S$\mu^2$ & \Area \\
transition			& (GHz)		& (K)	&  	&$\!\!\!\!\!\!\!\!$(Jy~km~s$^{-1}$) \\
\hline
 SO$_2$ &&&& \\
4$_{3,1}$-3$_{2,2}$			& 332.50524	& 15.34  & 6.92	 & 15.56\Mm0.64\\
8$_{2,6}$-7$_{1,7}$			& 334.67335	& 27.08  & 4.95	 & 13.91\Mm0.59\\
13$_{2,12}$-12$_{1,11}$$^b$	& 345.33854	& 76.41  & 13.41 & 13.82\Mm0.58 \\
16$_{4,12}$-16$_{3,13}$$^b$	& 346.52388	& 147.83 & 23.10 & 17.37\Mm1.15\\
19$_{1,19}$-18$_{0,18}$		& 346.65217	& 151.50 & 41.98 & 21.03\Mm0.58\\
21$_{2,20}$-21$_{1,21}$		& 332.09143	& 203.59 & 15.20 & 13.00\Mm0.61 \\
\hline
$^{34}$SO$_2$  &&&& \\ 
6$_{4,2}$-6$_{3,3}$		& 345.55309 	& 40.66  & 6.34  & 3.24\Mm0.60\\ 
7$_{4,4}$-7$_{3,5}$$^a$		& 345.51966 	& 47.09  & 8.09  & 1.44\Mm0.58 \\ 
8$_{4,4}$-8$_{3,5}$			& 345.16866 	& 54.46  & 9.77  & 4.43\Mm0.73 \\ 
9$_{4,6}$-9$_{3,7}$			& 345.28562 	& 62.72  & 11.41 & 5.20\Mm0.78\\ 
10$_{4,6}$-10$_{3,7}$		& 344.24535 	& 71.97  & 13.06 & 4.82\Mm0.66 \\ 
11$_{4,8}$-11$_{3,9}$		& 344.99816 	& 82.05  & 14.67 & 3.22\Mm0.69 \\ 
13$_{4,10}$-13$_{3,11}$		& 344.80791 	& 105.06 & 17.92 & 4.05\Mm0.67\\ 
15$_{4,12}$-15$_{3,13}$		& 344.98758 	& 131.76 & 21.20 & 4.92\Mm0.69 \\ 
16$_{4,12}$-16$_{3,13}$		& 332.83623	& 147.14 & 23.23 & 4.55\Mm0.62\\ 
19$_{1,19}$-18$_{0,18}$		& 344.58104 	& 151.10 & 42.24 & 4.91\Mm0.68 \\ 
17$_{4,14}$-17$_{3,15}$		& 345.92935	& 162.14 & 24.49 & 2.28\Mm0.57\\ 
\hline
OS$^{17}$O  &&&& \\ 
15$_{4,11}$-15$_{3,12}$$^a$& 345.93452	& 133.72 &125.43 	& 2.90\Mm0.53 \\
\hline
SO  &&&& \\
9$_8$-8$_7$	$^b$			& 346.52848	& 62.14  & 21.52 & 23.03\Mm1.15 \\
8$_8$-7$_7$				& 344.31061	& 70.96  & 18.56 & 15.56\Mm0.20 \\
\hline
$^{34}$SO &&&& \\
7$_8$-6$_7$ 				& 333.90098	& 63.84  & 16.24 & 10.01\Mm0.60\\
\hline
 CH$_3$OH &&&&  \\
2$_{2,1}$-3$_{1,2}$			& 335.13369 	& 28.59  & 0.31  &  3.12\Mm0.40 \\
7$_{1,7}$-6$_{1,6}$ A$+$		& 335.58200 	& 62.87  & 5.55  & 11.35\Mm0.44 \\
5$_{4,2}$-6$_{3,3}$			& 346.20278 	& 98.55  & 0.496 & 2.40\Mm0.54 \\
5$_{4,1}$-6$_{3,4}$			& 346.20437 	& 98.55  & 0.496 & 2.40\Mm0.54 \\
3$_{0,3}$-2$_{1,2}$ v$_t$=1	& 334.42659 	& 298.42 & 0.89  & 1.24\Mm0.44\\
16$_{1,15}$-15$_{2,14}$		& 345.90397 	& 316.05 & 7.13  & 6.03\Mm0.70 \\
18$_{2,16}$-17$_{3,14}$		& 344.10913 	& 402.88 & 5.31  & 4.13\Mm0.49 \\
19$_{1,19}$-18$_{2,16}$		& 344.44390 	& 434.70 & 6.00  & 3.26\Mm0.75  \\
18$_{-3,16}$-17$_{-4,14}$$^a$	& 345.91919 	& 442.83 & 5.61  & 2.02\Mm0.50\\
\hline
CO 3-2$^c$			& 345.79599	& 16.60  & 0.036 &  \\
HDCO 5$_{1,4}$-4$_{1,3}$$^a$& 335.09679	& 40.17  & 26.05 & 1.64\Mm0.40 \\
HC$_3 ^{15}$N 38-37		& 335.56088	& 297.98 & 526.9 & 2.08\Mm0.37 \\
HC$^{15}$N 4-3			& 344.20011 	& 24.78  & 35.65 & 4.76\Mm0.63 \\
H$^{13}$CN 4-3			& 345.33976	& 24.86  & 35.65 & $^b$\\
H$^{13}$CO$^+$ 4-3$^d$		& 346.99835	& 24.98  & 60.85 &-2.22\Mm0.51 \\
SiO 8-7					& 347.33082	& 58.35  & 76.79 & 2.58\Mm0.77 \\
\hline
\end{tabular}
\\
$^a$ Tentative detection. \\
$^b$ SO$_2$ 13$_{2,12}$-12$_{1,11}$ and H$^{13}$CN 4-3 are blended. The same holds for SO 9$_8$-8$_7$ and SO$_2$ 16$_{4,12}$-16$_{3,13}$. Therefore, their estimated line intensities are possibly not well determined and the quoted uncertainties come just from the fitting procedure. \\
$^c$ Very broad line, the cloud velocities are strongly affected by missing flux.
\\
$^d$ Tentative detection seen in absorption. \\
\end{center}
\end{table}%

\section{Results}\label{res}

\subsection{880~$\mu$m continuum emission}

Figure~\ref{Fcont} shows the 880~$\mu$m (340.0 GHz) continuum map toward GGD27 obtained with the SMA.  This map is in good agreement with  the subarcsecond 1.35 mm continuum maps by FL11a, where the sources were detected. Table~\ref{T1} shows the position, peak intensity and flux density of the sources. MM1 shows compact emission. A Gaussian fit to this source yielded a deconvolved  size of $\simeq0\farcs17$, indicating that the source radius is $\sim150$~AU, which is coincident with the upper limit found by FL11a.  However, the fit can not constrain whether the source is elongated in a particular direction. MM2(W) and MM2(E) are surrounded by weak emission at a $\sim$15~\mJy\ level extending $\sim1''$ (about 1700~AU). No dust continuum emission is detected around the warm molecular core \citep[WMC;][]{Qiu09}.

No continuum polarization is detected above 3-$\sigma$ level in the Stokes \stQ\ and \stU\ maps. This yields an upper limit (at 3-$\sigma$) of polarization of 0.8\% and 4.7\% for MM1 and MM2(W), respectively.  The MM1 upper limit is significantly lower than the typical values found in millimeter interferometric observations of dense cores around low and high mass star forming regions \citep[$\sim$2--10\%, e.g.][]{Girart99, LaiEtAl2003, Girart06, Girart09, Hull14, Zhang14}, but it is not far from the values found in disks around low--mass young stellar objects \citep{Hughes13, 2014Rao, Stephens14, Kataoka16}.

\subsection{Molecular emission in MM1}

Previous arcsecond observations show that in GGD27 there are three sites with bright molecular emission \citep[\eg][]{Qiu09}: two of them are sites of massive star formation, GGD27 MM1 and MM2, and the WMC, which does not have clear signs of star formation. The new subarcsecond angular resolution observations over the whole observed spectra (totaling about 7.8~GHz) show very different molecular characteristics in the three molecular sources (see Fig.~\ref{Fspec}).   MM2 and WMC present an almost empty spectra, showing no emission lines in the whole spectrum except for the CO 3--2 line, and marginal emission of the SO 9$_8$-8$_7$, CH$_3$OH 7$_{1,7}$-6$_{1,6}$ lines for WMC, and of the  \htcop\ 4-3 line for MM 2.  This is different to what has been found previously at both lower angular resolution and lower frequencies, specially toward WMC, where several relatively complex molecules were clearly detected, \eg\ CH$_3$OH, CH$_3$CN or HNCO \citep{Qiu09}.  MM2 exhibits strong CO emission associated with molecular outflows being powered by embedded protostars \citep{FL13}.

\begin{figure*}[h]
\includegraphics[width=17cm]{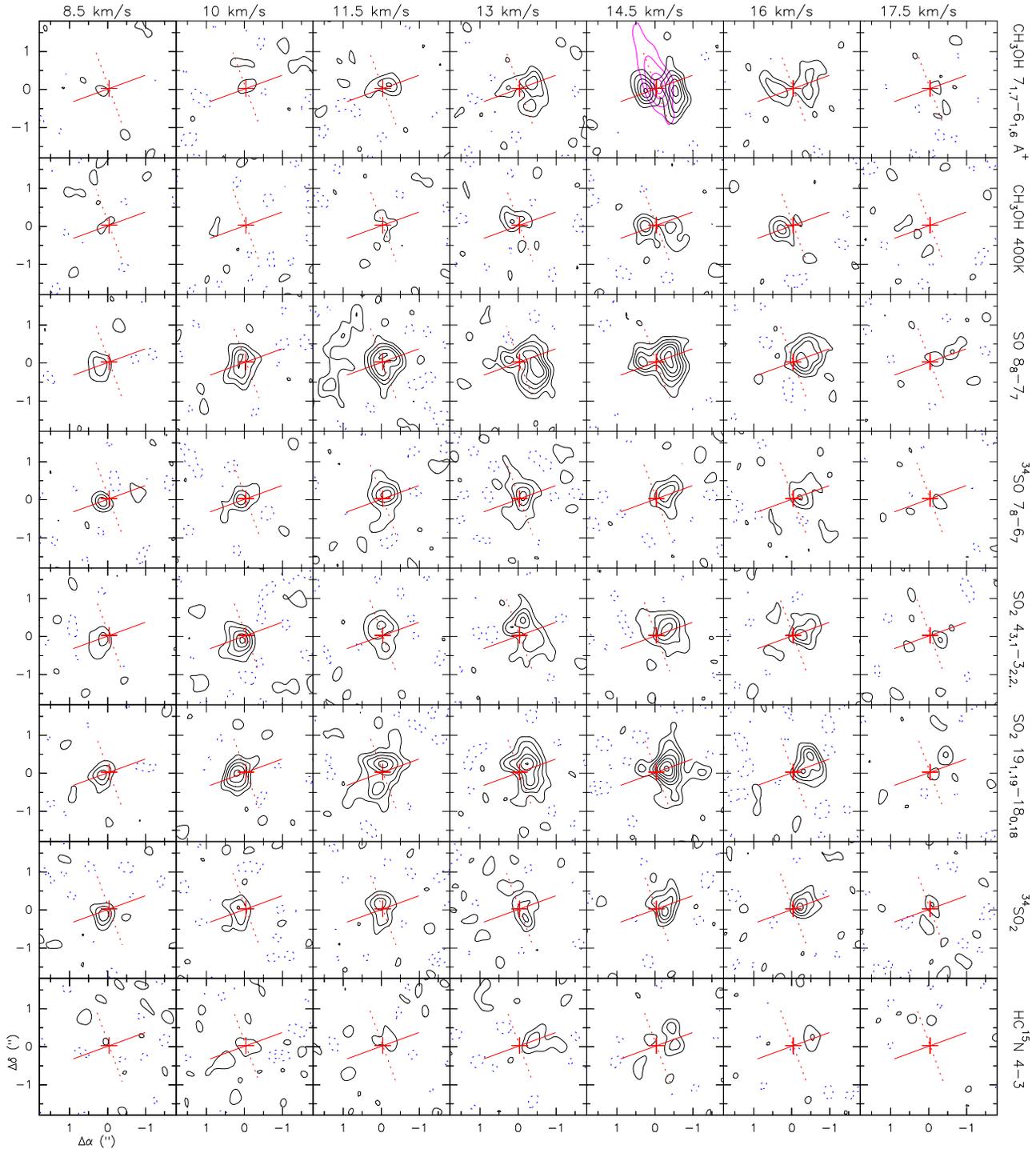}
\caption{Channel maps toward GGD27 MM1 of a representative set of molecular transitions. The molecular line transition is indicated in the right part of each row. \meta\ 400~K indicates the stacking averaged map of a combination of three methanol lines with an energy level of the lower state of $\sim 400$~K (16$_{1,15}$-15$_{2,14}$, 18$_{2,16}$-17$_{3,14}$ and 19$_{1,19}$-18$_{2,16}$).  Similarly, the \Tsod\  row is the result of stacking a combination of the detected \Tsod\ lines,  (except for the 6$_{4,2}$-6$_{3,3}$ and 7$_{4,4}$-7$_{3,5}$ lines). These lines have energies, for their lower levels, between 54 and 162~K.  The magenta contour emission in the top row shows the radio jet emission at 3.6 cm from \citep{Carrasco10}. Contours are multiple of 2--$\sigma$, with the first positive/negative contour at $\pm 2$--$\sigma$ level. The rms noise, $\sigma$, for each row of panels (from top to bottom): 64, 42, 70, 61, 74, 75, 23 and 65~\mJy. The red cross marks the position of GGD27 MM1 dust continuum source.  The red solid and dashed lines indicate the orientation of the disk and jet, respectively. 
}
\label{Fchan}
\end{figure*}

\begin{figure}[h]
\includegraphics[width=\columnwidth]{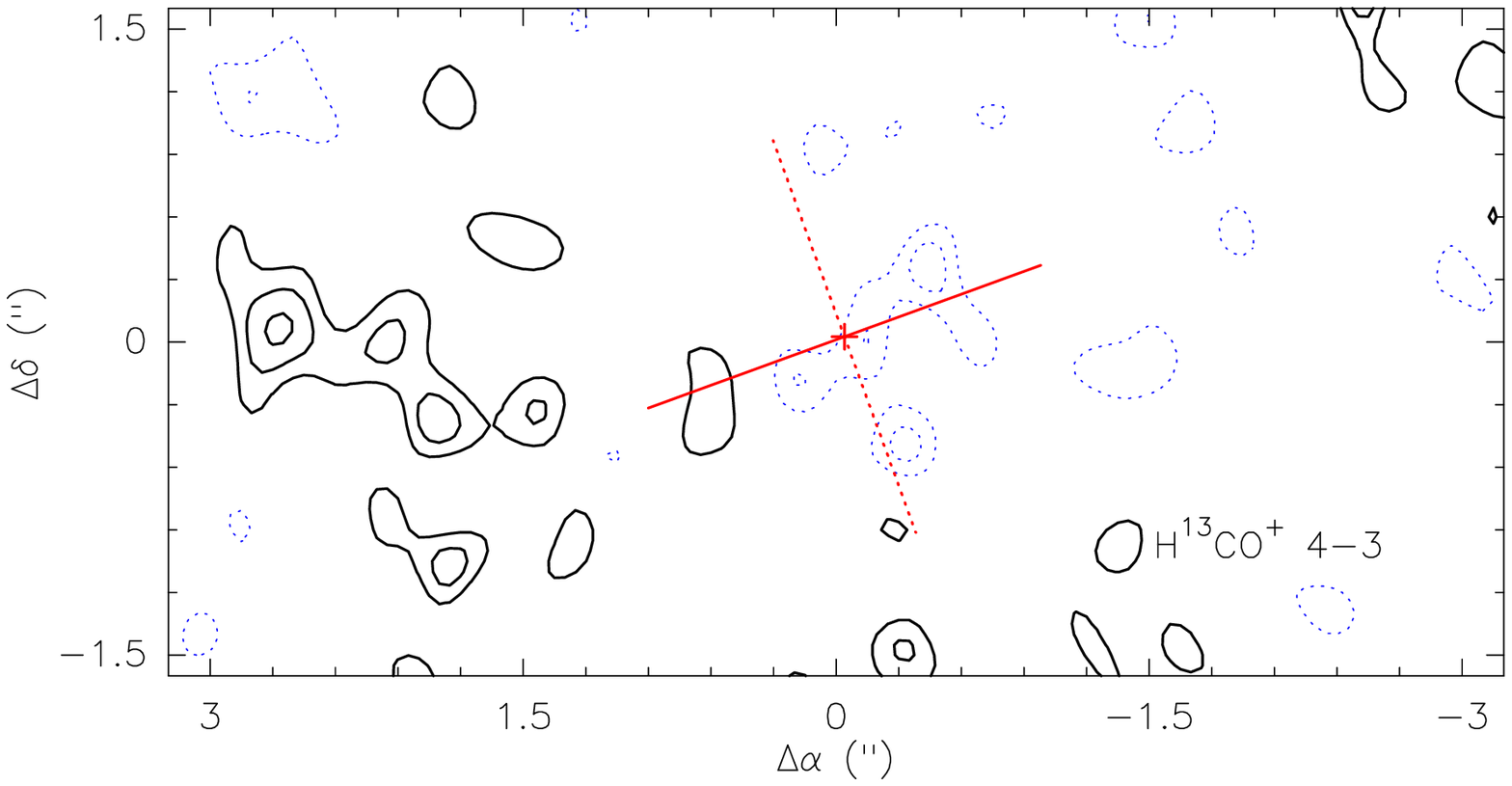}
\includegraphics[width=\columnwidth]{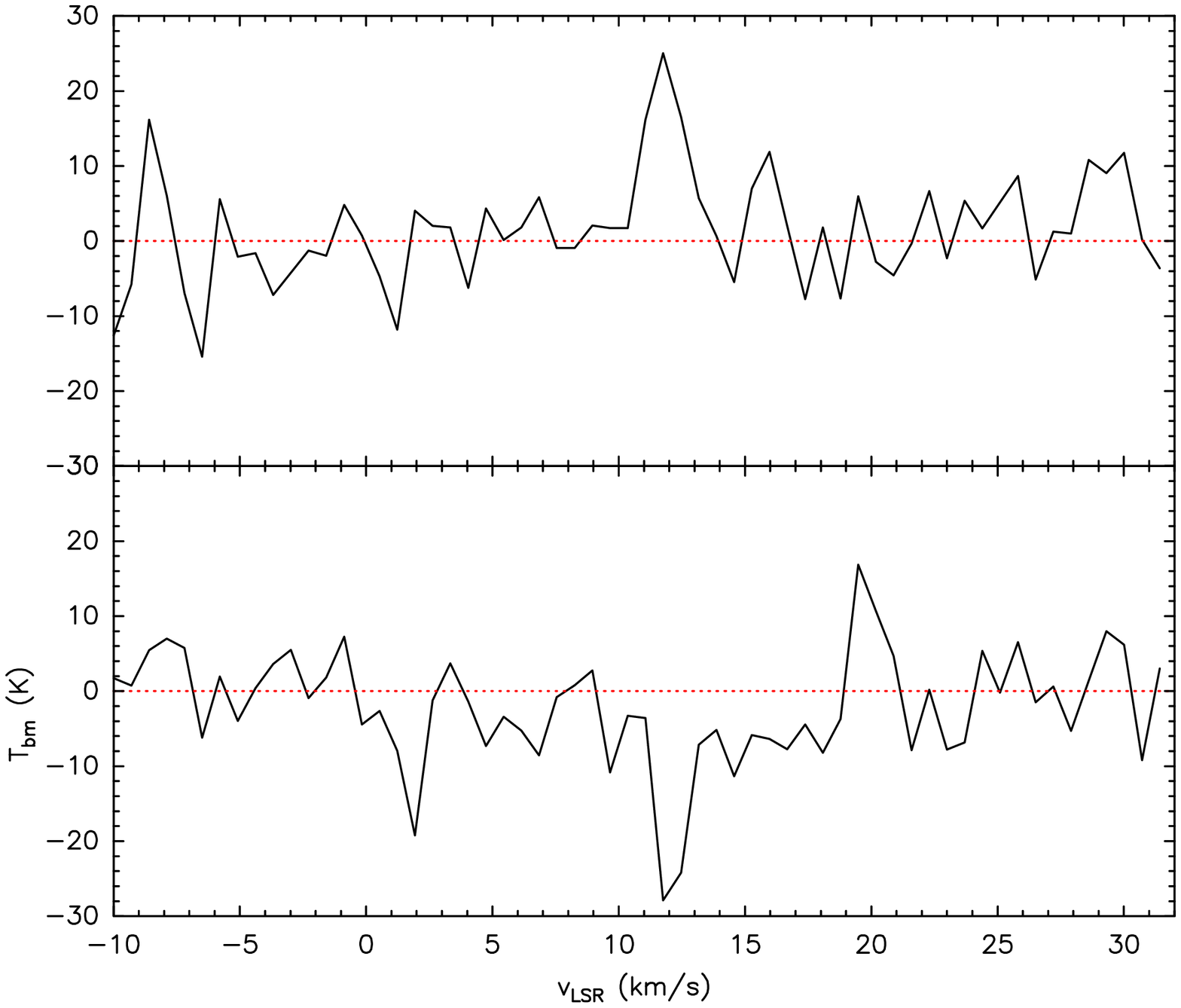}
\caption{
Top panel: Map of the  \htcop\ 4--3 at the \vlsr\ velocity of 11.5~\kms. Contours are $-3$, $-2$, 2, 3 and 4--$\sigma$, the rms noise of the map, 71~\mJy.
Two bottom panels:   \htcop\ 4--3 spectra toward GGD 27 MM1 (bottom panel) and toward a position located $-2\farcs7$   east of GGD 27 MM1 (middle panel).  The $rms$ noise of the spectra is 6.4~K.
}
\label{Fhtcop}
\end{figure}

In contrast to these two sources, GGD27 MM1 shows many molecular transitions along the observed bandwidth (see Figures~\ref{Fspec} and \ref{Fchan}). The spectra is dominated by SO$_2$, and SO isotopologues, as well as by \meta\ lines.  Other lines detected are isotopologues of  HCN, HC$_3$N and deuterated formaldehyde.  The \htcop\ 4--3 line is marginally detected in absorption towards the dust emission and in emission in an elongated structure east of the MM1 (the spectrum of these two features are shown in Fig.~\ref{Fhtcop}).  Table~\ref{T2} lists the parameters of the lines detected in MM1: line transition, frequency, line strength, energy level and integrated flux  of the detected molecular 
transitions.  The integrated flux was obtained by adding the emission for the channels where the emission is detected and over a region of $\sim2$~arcsec$^2$ on GGD27 MM1.

\begin{figure}[h]
\includegraphics[width=8cm]{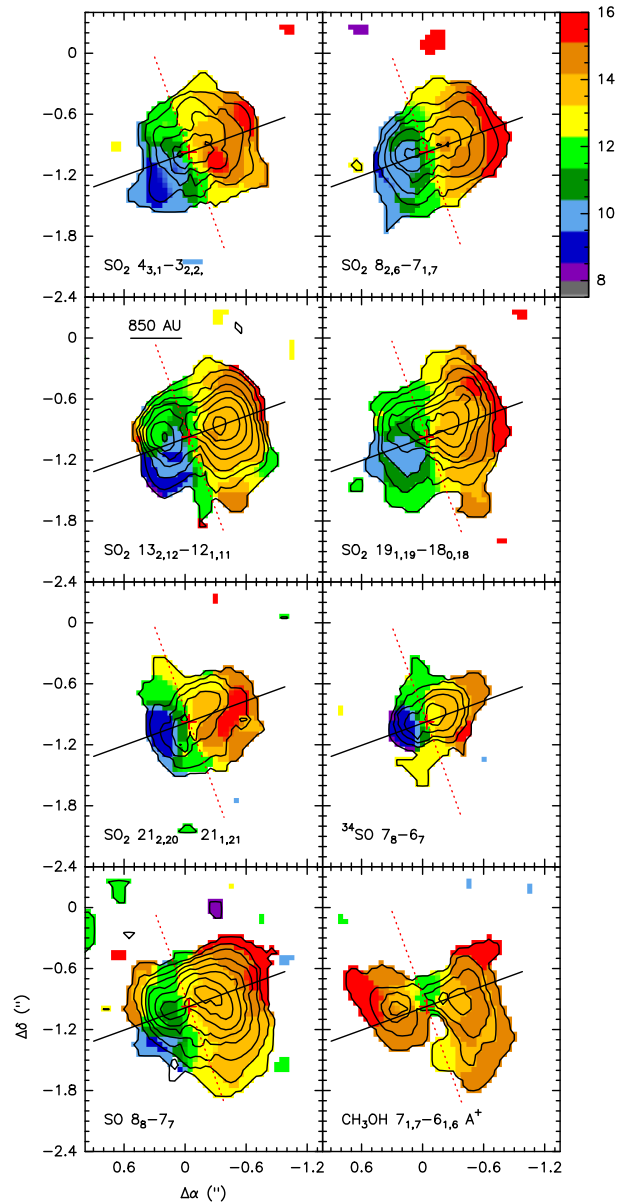}
\caption{Moment 0 (integrated intensity) images from different molecular lines toward GGD27 MM1 (contours) overlapping its moment 1 (centroid velocity) images (color scale). The red cross marks the position of GGD27 MM1 dust continuum source. The dashed red and solid black lines indicate the orientation of the jet and disk, respectively. Labels with the name of the lines are placed at the bottom of every panel. 
}
\label{Fmom}
\end{figure}

\begin{figure}[h]
\epsscale{1.0}
\plotone{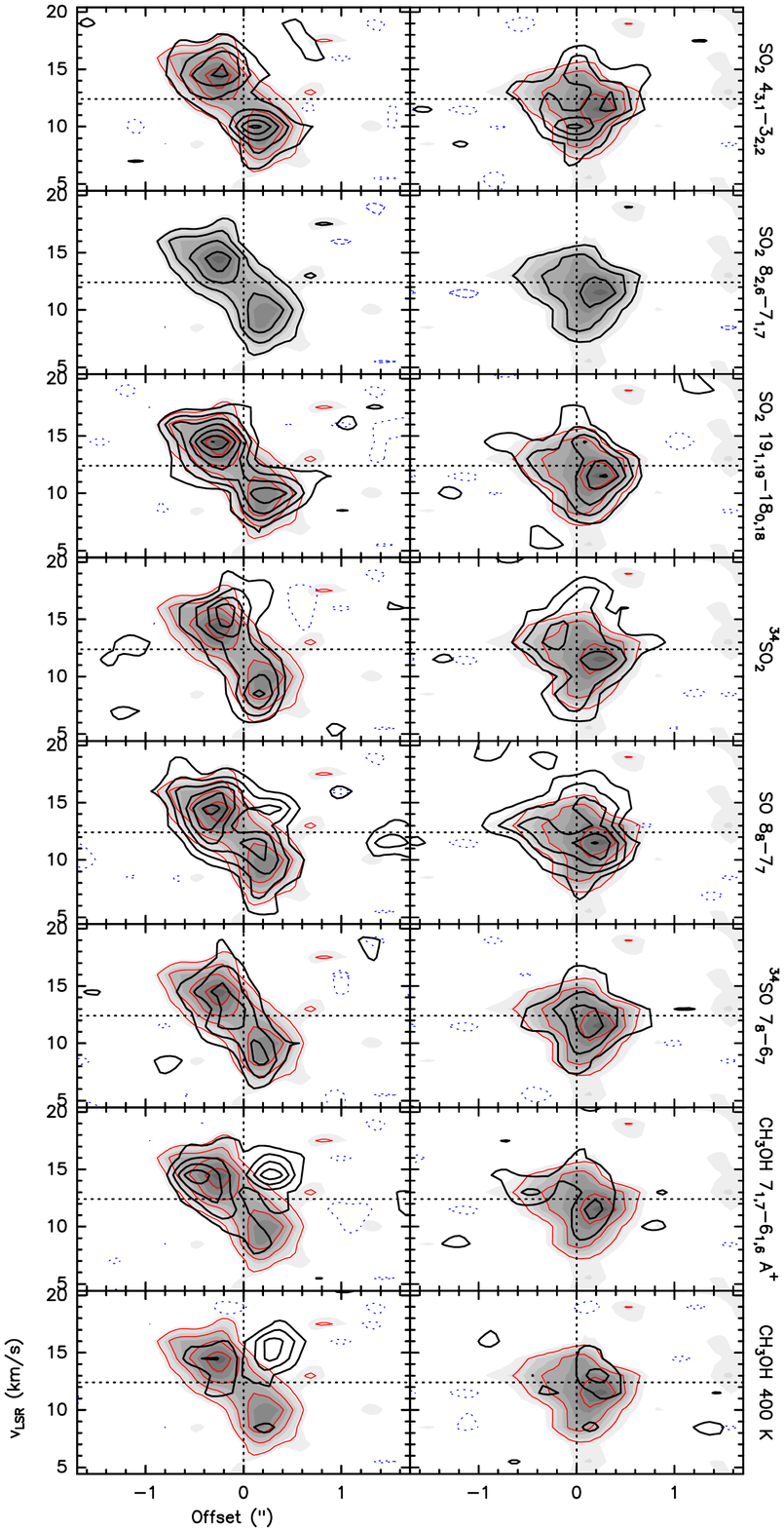}
\caption{Position--velocity cuts along the major axis of the GGD27 MM1 disk (left panels; $PA=111\arcdeg$) and minor axis (right panels; $PA=21\arcdeg$) for several lines (solid black contours and grey scale). For comparison, all panels are overlapped with the  \sod\ 8$_{2,6}$-7$_{1,7}$  line in red contours. Contours are in steps of 2--$\sigma$, starting at 2--$\sigma$.  The rms noise of the different lines are given in Fig.~\ref{Fchan} except for the \sod\ 8$_{2,6}$-7$_{1,7}$, which is  74~\mJy.
}
\label{Fpvobs}
\end{figure}

In order to better study and characterize the morphology and the kinematics of the detected molecular lines toward MM1, we present here channel maps of representative molecular lines (Fig.~\ref{Fchan}), the zero and first order moment maps (integrated emission and velocity field weighted by intensity, respectively; Fig.~\ref{Fmom}), and the position--velocity cuts along the major and minor axis of the molecular emission ($PA=111\arcdeg$ and 21\arcdeg, respectively;  Fig.~\ref{Fpvobs}) .  For three high-excitation \meta\ lines (16$_{1,15}$-15$_{2,14}$, 18$_{2,16}$-17$_{3,14}$ and 19$_{1,19}$-18$_{2,16}$) and most of the detected \Tso\ lines  (see the caption of Fig.~\ref{Fchan} for more details), the maps were combined in order to increase their signal-to-noise ratio.  The three figures show that there is a distinctive velocity gradient along the major axis of the molecular disk-like structure (blueshifted and redshifted gas to the east and west, respectively).  This velocity gradient is along the major axis of the molecular emission and perpendicular to the HH 80-81-80N radio jet.  These properties are more clearly seen in the  SO$_2$, \Tsod\ and \Tso\ lines. Indeed, the first order moment maps of these lines are in agreement with those previously reported (FL11b, CG12), but the maps presented here resolve significantly better the velocity gradient. The emission along the major axis extends $\sim1\farcs4$ or 2400~AU. The position-velocity cuts along the major axis also show a clear linear velocity gradient. The different lines of the \sod\ main isotopologues show the same gradient within the uncertainties. However, the combination of \Tsod\ lines show a steeper gradient and more compact emission. The peak emission appears also at higher velocities. The steeper velocity gradient is also seen in the \Tso\ line. The position-velocity cuts along the minor axis present a broaden of the velocity range at the center for all the cases. All these features suggest that the emission traces a rotating molecular disk around the massive star, as previously suggested in FL11b and CG12.

The methanol emission appears to partially depart from the rotating disk pattern, which is clearly shown in the integrated emission and the velocity field of the \meta\ 7$_{1,7}$-6$_{1,6}$ A$^+$  (Fig.~\ref{Fmom}). The position-velocity cut along the major axis shows that whereas part of the emission arises from the rotating disk, there is a clear redshifted knot (peaking at \vlsr\ =16.0~\kms) in the blueshifted side of the rotating disk (see Fig.~\ref{Fpvobs}). Furthermore, the central channels of the \meta\ 7$_{1,7}$-6$_{1,6}$ A$^+$ line shows that the emission is elongated along the jet axis (\vlsr\ = 14.0 and 16.0~\kms\ channels in Fig.~\ref{Fchan}).   This suggests that not all the methanol emission is associated with the rotating disk, so some may arise from the walls of the cavity excavated by the outflow. The SO 8$_8$-7$_7$ line seems to be tracing both features, the rotating disk and the walls of the cavity.  Finally, 
the HC$^{15}$N 4--3 line does not clearly show the velocity gradient (see Fig.~\ref{Fchan}), although this could be due to the low signal-to-noise ratio of the emission of this line.



\section{Analysis: The thin-disk model for the SO$_2$ and SO emission}
\label{ana}

The SO$_2$ and SO lines appear to show a clear velocity gradient 
along the major axis of the disk, suggestive of rotation, as observed
previously at lower angular resolution (FL11b). 
To better constrain the kinematics of the disk, we modeled the emission with a
rotating, geometrically thin disk,  
using the SO$_2$ transitions
4$_{3,1}$-3$_{2,2}$,			    
8$_{2,6}$-7$_{1,7}$,			    
19$_{1,19}$-18$_{0,18}$, and	
21$_{2,20}$-21$_{1,21}$;		  
the $^{34}$SO 
7$_8$-6$_7$ transition;			  
and a sum of 				          
$^{34}$SO$_2$ transitions for improving the signal-to-noise ratio.

\begin{figure}[h]
\epsscale{1.0}
\plotone{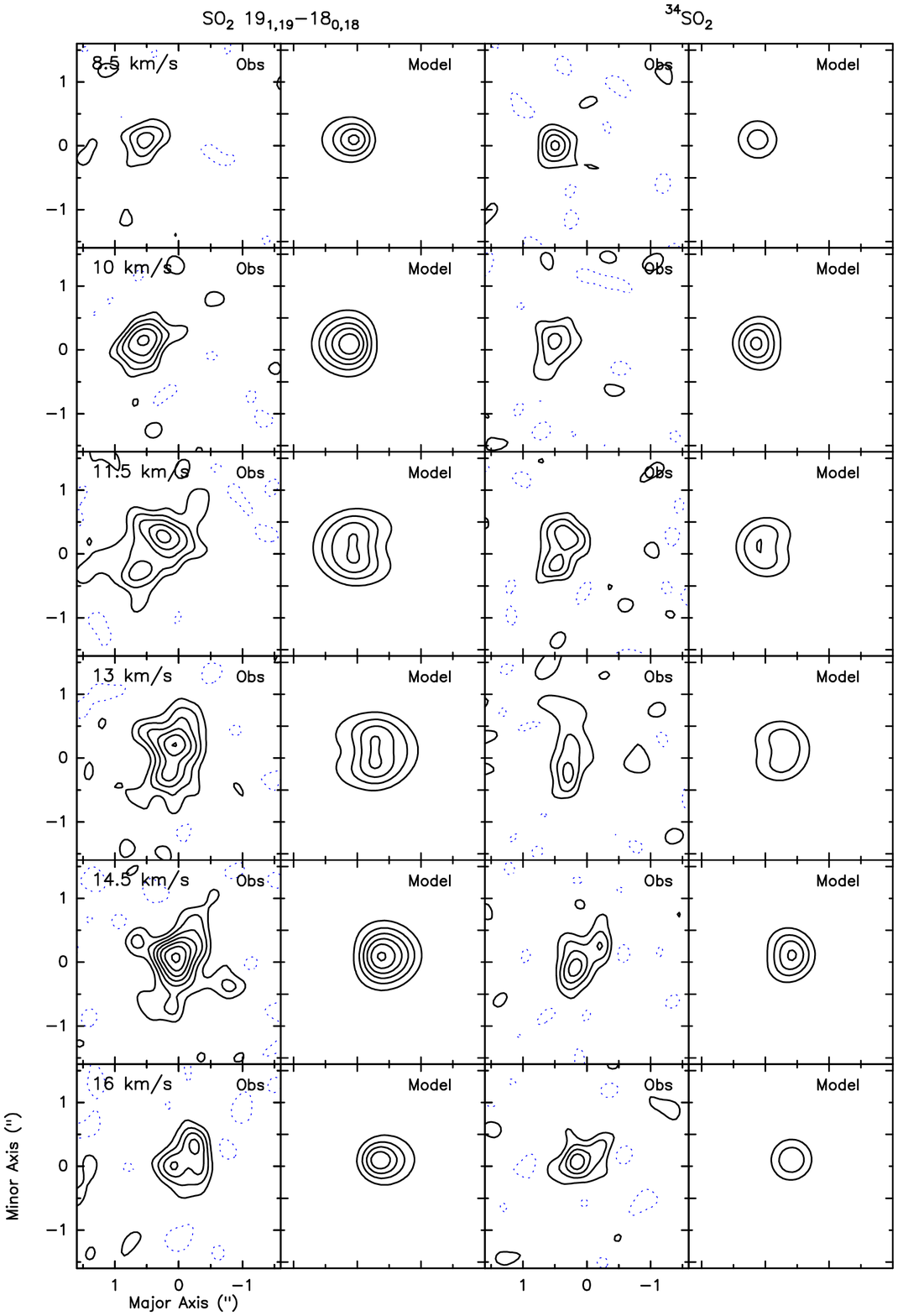}
\caption{Channel maps of the SMA SO$_2$ 19$_{1,19}$-18$_{0,18}$ and the combined \Tsod\ lines (first and third column, respectively). Best thin-disk models for these two lines are shown in the second and fourth  columns. These maps have been rotated $21\arcdeg$ so the major and minor axes are the x and y axes, respectively.}
\label{Fmodel}
\end{figure}

\begin{figure}[h]
\includegraphics[width=\columnwidth]{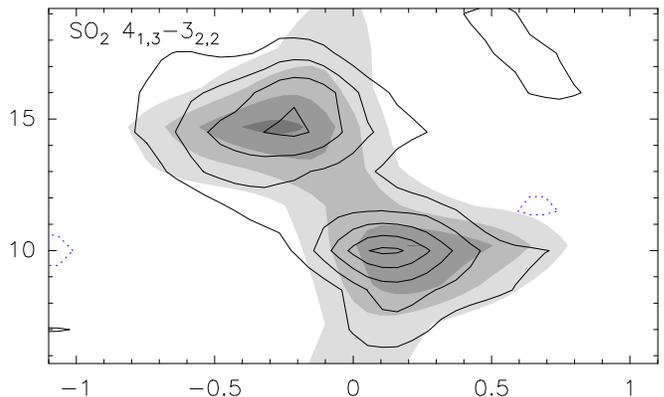}
\caption{
Position-velocity cut along the major axis for the SO$_2$ 4$_{3,1}$-3$_{2,2}$ line (contour map) and the best disk model for this transition (grey scales). Contours  are in steps of 2-$\sigma$, starting at 2-$\sigma$.
} 
\label{Fpv-model}
\end{figure}

\begin{table}[htb]
\centering
\caption{Thin disk model: 
Weighted average of best fit parameters for all transitions
\label{Tfit1}}
\begin{tabular}{llc}
\hline
Parameter & 
\multicolumn{1}{c}{Units}  &
\multicolumn{1}{c}{Value}  \\
\hline
Linewidth   $\Delta v$                     &(km s$^{-1}$) &\phs$3.3\pm0.3$ \\
Disk center $x_0$                          &(arcsec)      &$-0.13\pm0.05$ \\
Disk center $y_0$                          &(arcsec)      &$-0.40\pm0.04$ \\
Disk central $v_0$                         &(km s$^{-1}$) &\phs$12.4\pm0.2$\phn\\
Rotation vel. $v_r\sin i$ \tablenotemark{a}&(km s$^{-1}$) &$-2.5\pm0.5$ \\
Disk inclination $i$                       &(degrees)     &\phs$47\pm8$\phn \\
\hline
\end{tabular}
\\
$^a$ At the reference radius $r_0=1''$.
\end{table}

\begin{table}[htb]
\centering
\caption{Thin disk model: Disk inner and outer radii fitted for every transition
\label{Tfit2}}
\begin{tabular}{lcc}
\hline
&
Inner Radius &
Outer Radius\\
Transition & 
(arcsec) &
(arcsec)\\
\hline
SO$_2$ 4$_{3,1}$-3$_{2,2}$      & $0.04\pm0.01$  & $0.74\pm0.01$ \\
SO$_2$ 8$_{2,6}$-7$_{1,7}$      & $0.04\pm0.01$  & $0.75\pm0.01$ \\
SO$_2$ 19$_{1,19}$-18$_{0,18}$  & $0.04\pm0.01$  & $0.82\pm0.01$ \\
SO$_2$ 21$_{2,20}$-21$_{1,21}$  & $0.09\pm0.01$  & $0.69\pm0.01$ \\
$^{34}$SO 7$_8$-6$_7$           & $0.03\pm0.01$  & $0.54\pm0.01$ \\
$^{34}$SO$_2$                   & $0.00\pm0.01$  & $0.59\pm0.01$ \\
\hline
\end{tabular}
\end{table}%

The position angle of the projection of the disk axis on the plane of the sky
(minor axis position angle) was fixed to $21\arcdeg$, which is the position
angle of the radio jet \citep{Marti93}.
The inclination of the disk, $i$, was defined as the angle between the disk
axis and the line of sight ($i=0^\circ$ for a face-on disk).  
We considered a rotation velocity given by a power law of the radius,   
$v_r (r/r_0)^{q_r}$, where $r_0$ is an arbitrary reference radius ($r_0=1''$)
and
$v_r$ is the rotation velocity at the reference radius. 
We assumed that the molecular emission arises from an area of the disk between
an inner radius $r_{i}$ and an outer radius $r_{o}$.

We computed, for each point of a regular grid on the plane of the sky, the
projection of the rotation velocity of the corresponding point of the disk along
the line of sight $v_z$. A Gaussian line profile of width $\Delta v$ and
centered on $v_z$ was added to the channels associated with the grid point. The
intensity of the Gaussian was taken to follow a power-law dependence on the disk
radius, $r^{q_d}$.
Finally, each channel map was convolved spatially with a Gaussian beam of width
$\Delta s$. However, the intensity scale of the channel maps is arbitrary. A
scaling factor, the same for all channel maps, was obtained by minimizing the 
sum of the squared differences between the data channel maps  and the synthetic 
channel maps.
The model depends on a total of 11 parameters,  namely
the beamwidth, $\Delta s$;
the linewidth, $\Delta v$;
the disk center position, $(x_0, y_0)$; 
the disk systemic velocity, $v_0$; 
the disk inner and outer radii, $r_i$ and $r_o$; 
the radial dependence power-law index of the intensity, $q_d$; 
the projection of the disk rotation velocity at the reference radius, 
$v_r \sin i$;  
the radial dependence power-law index of the rotation velocity, $q_r$; 
and 
the disk inclination angle, $i$. 

Some of the parameters are known beforehand, such as $\Delta s$, the synthesized
beamwidth for every transition. Some other were taken as fixed: the rotation
power-law index was taken as that of a Keplerian rotation, $q_r=-0.5$; and the
intensity power-law index, which takes into account the radial dependence of
density and temperature, was taken as $q_d=-1$. 

The fitting procedure was the sampling of the multidimensional parameter space, 
using the same procedure as that described in \citet{Girart14}  and 
\citet{Estalella16}.  
The parameter space was searched for the minimum value of the rms fit
residual.   Once a minimum of the rms fit residual was found, the uncertainty in
the parameters  fitted was found as the increment of each of the parameters of
the fit necessary to  increase the rms fit residual a factor of
$[1+\Delta(m, \alpha)/(n-m)]^{1/2}$, where
$n$ is the number of data points fitted, 
$m$ is the number of parameters fitted, and 
$\Delta(m, \alpha)$ is the value of $\chi^2$ for $m$ degrees of freedom (the
number  of free parameters) and $\alpha$ is the significance level
($\alpha=0.68$ for 1-$\sigma$ uncertainties). 

The fit was performed in two steps.
In a first run, the 8 free parameters
($\Delta v$, $x_0$, $y_0$, $v_0$, $r_i$, $r_o$, $v_r\sin i$, and $i$) were
fitted simultaneously for the 6 transitions. From this run we obtained the
weighted average of the best fit values for the parameters that do not
depend on the transition, $\Delta v$, $x_0$, $y_0$, $v_0$,
$v_r\sin i$, and $i$ (see Table \ref{Tfit1}). The value obtained for the disk
inclination was $i=47^\circ\pm8^\circ$. The projection of the rotation velocity
at $r_0=1''$ obtained is $v_r\sin i=-2.5\pm0.5$\kms, corresponding to a deprojected
rotation velocity $v_r=-3.4\pm0.8$\kms. 

In a second run we set as constant the 8 parameters obtained from the first fit
and we fitted the inner and outer radii of the disk for every transition. The
results obtained are shown in Table \ref{Tfit2}. The inner radius obtained for
all the transitions was $r_i<0\farcs1$. The outer radius for the SO$_2$
transitions is $r_o\simeq0\farcs75$ while for the $^{34}$SO and $^{34}$SO$_2$
transitions it is smaller, $r_o\simeq0\farcs55$. 
Fig.\ \ref{Fmodel} shows, as an example of the obtained fits, the channel maps  for the SO$_2$ 4$_{3,1}$-3$_{2,2}$ and 19$_{1,19}$-18$_{0,18}$ transition data and their best fit model. Fig.\ \ref{Fpv-model} shows the position-velocity cut of the SO$_2$ 4$_{3,1}$-3$_{2,2}$  line and the best fit model along the major axis.

\section{Discussion}\label{dis}

\subsection{Rotating disk-like structure and the stellar mass}

The analysis of the velocity field derived from the \sod\ and SO show that the emission is more compact in their $^{34}$S isotopologues and in the transition of the main \sod\ isotopologue with the highest energy level (20$_{2,20}$-21$_{1,21}$).  This is a consequence of the increase of the column density and temperature  toward the center of the disk.  However, the analysis of the velocity field derived from the \sod\ and SO isotopologues is limited by the signal-to-noise of the data (the highest value is $\simeq 15$). This precludes the detection of the faint gas at higher velocities, which is expected to arise closer to the protostar (according to the model, this gas has intensities below the detectability of our  data). Thus, our data cannot discern whether the rotation is Keplerian. However, the dynamical mass of the system (protostar and disk) can be constrained by using two approximations. First we assume that the velocity in the whole disk is Keplerian ($M=v^2\cdot r/G$). Taking the obtained rotation velocity (3.4\kms) at the reference radius ($1\arcsec\simeq1700$~AU) we estimate a dynamical mass of $22\pm4$\msun. In second place, we assume that the gas is not all in Keplerian motion but still gravitationally bound. In this case, we can still balance the gravitational and the kinetic energy of the gas at the outer edge of the disk ($r_o\sim0\farcs75\simeq1300$~AU and $v_{rot}=\Delta v/(2\sin{i})\simeq2.3$\kms) to derive a dynamical mass ($M=v^2\cdot r/(2G)$) of $8\pm2$\msun. With the present data we cannot discern whether the Keplerian approximation stands for all the observed gas surrounding MM1 or there is a centrifugal barrier separating a rotating infalling region (outer disk/envelope) from a Keplerian rotating region (inner disk), as found in some lower mass protostars \citep{Sakai14}. In any case, the latter value of 8\msun is similar to that derived previously in (FL11b) where the virial assumption was used as well. From the dust emission at 1.36~mm, and assuming optically thin emission, the mass from the disk is $\simeq4$~\msun\  (FL11a). This suggests that the stellar mass should be between 4 and 18~\msun. Therefore, the stellar mass is loosely constrain from the velocity analysis. The bolometric luminosity of this source is also not well constrained, but should be between 3300~\lsun and $2 \times 10^4$~\lsun (FL11b). This is because there is contamination from other sources at mid to far infrared wavelengths \citep[e.g.,][]{Aspin91, Aspin92, Qiu08} and, in particular, from  the nearby massive MM2  (FL11b).  Yet, the range of values for masses is compatible with the possible values of the bolometric luminosity because a significant fraction of the luminosity comes from accretion  \citep{Yorke02, Hosokawa10}.

\subsection{Chemical composition of the MM1 rotating disk-like structure}

Despite of having an observed rich spectrum, MM1 has a less rich  chemistry than standard hot cores.  The spectrum is dominated by sulfurated molecules and the lines that more clearly trace the rotating disk are \sod\ and SO isotopologues, but also H$_2$CO (FL11b). Previous observations also show that  CH$_3$CN, OCS and HNCO are also bright in MM1 \citep{Qiu09}, but the limited angular resolution of these data prevents from confirming whether their emission arises from the rotating disk.

\citet{Girart13} observed almost the same frequency range as the one presented here toward the DR 21(OH) massive star-forming region. The DR 21(OH) SMA 6 and SMA 7 submillimeter sources, known to be very dense hot cores, show the same lines detected in MM1. In these two hot cores there are other molecular species that show emission as bright as the \Tsod\ lines, specially methyl formate (CH$_3$OCHO), but also  dimethyl ether (CH$_3$OCH$_3$), formic acid (HCOOH), nitrogen monosulfide (NS), methanimine (CH$_2$NH) and ethyl cyanide (CH$_3$CH$_2$CN). None of these molecules are detected toward MM1. The molecular emission associated with MM1 arises from hot, 120--160~K, and very dense gas (FL11b), excluding the possibility that the difference between DR21(OH) and MM1 is due to a different excitation conditions. Thus,  the lack of complex molecules in MM1 indicates a chemical differentiation with respect to the DR 21(OH) hot cores.  

Recent observations toward disks around protostars show that the SO molecule is an excellent probe of the shocks generated at the position of the centrifugal barrier \citep{Sakai14, Oya16}.  The centrifugal barrier is the transition region between the envelope and the rotationally supported disk, where most of the kinetic energy of the infalling envelope is converted into rotational energy \citep{Sakai14,Sakai16}. Furthermore, chemical models predicts that not only SO but also \sod\ are enhanced in shocks with moderate shock velocities (few \kms) generated in dense molecular environments, $n$({H$_2$)$\sim 10^4$-$10^6$~\cmt,  \citep{Forets93}.  Indeed, these two species are enhanced in shocks associated with  molecular outflows \citep{Bachiller97, Codella03, Podio15}.

Alternatively, the apparently sulfurated dominated chemistry of the rotating disk-like structure around MM1 could be a consequence of MM1 being more evolved than the typical massive hot molecular cores \citep[e.g.,][]{Charnley97, Minh16}. Indeed, a chemical study of different species found that most molecules that are abundant in the hot molecular phase have their abundances decreased in UCHII phase, but this decrease is not so significant in some sulfurated molecules such as SO and SO$_2$ \citep[e.g.,][]{Gerner14}.  In the case of SO$_2$, the UV photodissociation of the water, yielding OH,  can enhance this molecule through the SO+OH$\rightarrow$SO$_2$ reaction \citep{Charnley97, Tappe08, Qiu09}. In MM1, the source of UV radiation may arise from the strong shocks generated in the powerful jet \citep{Carrasco10},  which would illuminate the surface of the molecular disk.  A comparison between the spectral energy distribution of  four massive cores without HII regions found that GGD 27 MM1 appears to be the most evolved one \citep{Herpin09}. We speculate that the spectral features of GGD 27 MM1 indicate that this source is in the verge of forming an ultracompact (UC) HII region.

\subsection{The outflow cavity walls }

As noted in Section \ref{res},  the spatial distribution of the \meta\ emission partially departs from the expected pattern for the rotating molecular disk.  This is better seen in the first order momentum map (Fig. \ref{Fmom}) and in the position--velocity cuts (Fig. \ref{Fpvobs}).   In this latter figure there is an emission feature at \vlsr\ = 14.5 \kms\ just east of the disk center that does not follow the rotation pattern. The clump appears in the northern side of the blueshifted side of the rotating disk-like structure. This clump is also clearly seen in the SO 8$_8$-7$_7$ emission.  This feature along with the overall morphology of the \meta, with the emission extended along the radio jet direction, suggests that the emission from \meta\ is tracing the outflow cavity walls  \citep[the outflow is probably atomic and ionized since no high velocity CO emission is detected at these scales, see][]{FL13}. The SO emission, along with the H$_2$CO 3$_{1,2}$-2$_{1,1}$ line emission  (FL11a), appears to arise from both the rotating disk and the outflow cavity walls.

\subsection{The lack of molecular emission associated with MM2 and WMC}

The lack of molecular emission associated with MM2 and WMC may seem surprising, especially when they have bright 1 mm molecular emission  at an angular resolution of $\simeq3''$ \citep{Qiu09}.  These observations showed that WMC and, in a less extent, MM2  are prominent in lines of \meta, CH$_3$CN, H$_2^{13}$CO, SO, OCS and HNCO with upper energy levels of $\la 100$~K. In contrast, none of the higher excitation lines detected in MM1 are detected in these two sources \citep[][FL11a]{Qiu09}.  This is because of the lower temperature ($\sim 40$~K) of these two sources. Thus, given the high volume density of MM2,  $\sim 10^9$~\cmt, one should expect to detect the SO and \meta\ lines with similar energy levels and line strength as those detected at 1~mm, such as the \meta\ 7$_{1,7}$-6$_{1,6}$ A$^+$ transition (see Table~\ref{T2}).   But these are barely detected only in WMC. The rms in brightness temperature of our observation is about 4--5 K, so we are sensitive (at 3--$\sigma$)  to line intensities of $\ga$12-15 K.  This indicates that the molecular gas around these two sources is extended and, in addition, since the upper limits for the brightness temperature are significantly lower than the gas kinetic temperature, the observed molecular lines are optically thin. 

The SO$_2$ lines represent a different scenario, since they have not been detected toward WMC and MM2 at lower angular resolution and frequency \citep[][FL11a]{Qiu09}.  In this case, \citet{Qiu09} interpreted this as a chemical differentiation because of their different evolutionary stages. MM2 has a similar mass as MM1 but it has a lower average volume density because it is more extended (FL11a). This is an indication that MM2 is possibly younger, suggesting that the accreted mass to this young stellar object is still smaller or that the accretion rate is not as high as in MM1. The WMC could be a less massive object than MM1 and MM2 because of the non-detection of the dust continuum emission.

\section{Conclusions}\label{con}
We present  subarcsecond angular resolution observations carried out with the Submillimeter Array (SMA) at 880~$\mu$m toward the GGD27 system. We do not detect any polarized continuum emission and place upper limits for the polarized intensity on GGD27~MM1 ($\lesssim0.8\%$) and GGD27~MM2 ($\lesssim4.7\%$).  The MM1 spectrum is dominated by the presence of several sulfur-bearing species tracing the disk-like structure (SO and SO$_2$ isotopologues mainly), but it also shows HCN-bearing and \meta\ lines. The \meta\ and SO emission is in part tracing the disk rotation, but also comes from the cavity walls of the outflow excavated by the jet. \htcop\ is seen in absorption against the MM1 continuum emission.  We discuss that the abundance of sulfurated lines in the spectrum (which is not common in other massive hot cores) could indicate the presence of shocked gas at the disk's centrifugal barrier or that MM1 is a hot core at an evolved stage. 

The $0\farcs4$ SMA resolution allows to clearly resolve the SO$_2$ disk's kinematics.  MM1 shows a velocity gradient along the disk's major axis, which is steeper for the \Tsod\ and \Tso\ lines, since their emission is more compact than for the SO$_2$ lines. We make a fit to the SMA observations using a thin-disk Keplerian model which fully characterize the geometry of the disk.  The results of our fit give an inclination of 47$\degr$, an inner radius of $\lesssim0\farcs1$ ($\lesssim170$~AU) and an outer radius between $0\farcs55$ (950~AU) and $0\farcs75$ (1300~AU), depending on the molecular tracer.  We also constrain the radial velocity at a putative radius of 1700~AU in 3.4\kms. Using these results, the estimated protostellar dynamical mass of MM1 is 4-18\msun. 

Finally, we also detect molecular emission from the MM2 and WMC cores (MM2 is also resolved in continuum emission). They show in comparison with MM1 an almost empty spectra suggesting that they are associated with extended emission (detected in previous low-angular resolution observations), and therefore indicating the presence of a very young system (MM2), still in a very early stage of accretion, or/and the presence of a less massive object (WMC).

\acknowledgments
We thank all members of the SMA staff that made these observations possible. 
The Submillimeter Array is a joint project between the Smithsonian Astrophysical Observatory and the Academia Sinica Institute of Astronomy and Astrophysics, and is funded by the Smithsonian Institution and the Academia Sinica. 
JMG, RE, GB and CJ are supported by the MINECO (Spain) AYA2014-57369-C3- grant. SC acknowledges support from DGAPA, UNAM and CONACyT, M\'exico.

\end{document}